\documentclass[9pt,shortpaper,twoside,web]{ieeecolor}
\usepackage{generic}
\usepackage{cite}
\usepackage{amsmath,amssymb,amsfonts}
\usepackage{algorithmic}
\usepackage{graphicx}
\usepackage{textcomp}
\usepackage{makecell}
\usepackage{float}
\floatstyle{ruled}
\newfloat{algorithm}{thp}{lop}
\floatname{algorithm}{Algorithm}

\def\BibTeX{{\rm B\kern-.05em{\sc i\kern-.025em b}\kern-.08em
    T\kern-.1667em\lower.7ex\hbox{E}\kern-.125emX}}
\begin{document}
\title{Maximum entropy based non-negative optoacoustic \\tomographic image reconstruction}
\author{Jaya Prakash$^{\dagger}$, Subhamoy Mandal$^{\dagger}$, \IEEEmembership{Member, IEEE}, Daniel Razansky,  \IEEEmembership{Member, IEEE}, \\and Vasilis Ntziachristos, \IEEEmembership{Senior Member, IEEE} 
\thanks{This article has been accepted for publication in IEEE
Transactions on Biomedical Engineering. DOI: 10.1109/TBME.2019.2892842}
\thanks{$^{\dagger}$J.P. and S.M. contributed equally to this work.}
\thanks{J.P. acknowledges support from the Alexander von Humboldt Postdoctoral Fellowship Program. S.M. acknowledges support from DAAD PhD Scholarship Award (A/11/75907) and IEEE Richard E. Merwin Scholarship. D.R. acknowledges funding support from the European Research Council (ERC-2015-CoG-682379), US National Institutes of Health (R21-EY026382-01), Human Frontier Science Program (RGY0070/2016) and Deutsche Forschungsgemeinschaft (RA1848/5-1). V.N. acknowledges funding support from European Research Council (694968, ERC-PREMSOT).}
\thanks{J.P., S.M., D.R., and V.N. are with the Institute of Biological and Medical Imaging, Helmholtz Zentrum Munich, Ingolstaedter Landstr. 1, D-85764 Neuherberg, Germany, and also with the Chair for Biological Imaging, Technical University Munich, Ismaningerstr 22, D-81675, Munich, Germany. (e-mail: v.ntziachristos@tum.de).}}

\maketitle

\begin{abstract}
\textbf{Objective:} Optoacoustic (photoacoustic) tomography is aimed at reconstructing maps of the initial pressure rise induced by the absorption of light pulses in tissue. In practice, due to inaccurate assumptions in the forward model, noise and other experimental factors, the images are often afflicted by artifacts, occasionally manifested as negative values. The aim of the work is to develop an inversion method which reduces the occurrence of negative values and improves the quantitative performance of optoacoustic imaging. 
\textbf{Methods:} We present a novel method for optoacoustic tomography based on an entropy maximization algorithm, which uses logarithmic regularization for attaining non-negative reconstructions. The reconstruction image quality is further improved using structural prior based fluence correction. 
\textbf{Results:} We report the performance achieved by the entropy maximization scheme on numerical simulation, experimental phantoms and \textit{in-vivo} samples. 
\textbf{Conclusion:} The proposed algorithm demonstrates superior reconstruction performance by delivering non-negative pixel values with no visible distortion of anatomical structures. \textbf{Significance:} Our method can enable quantitative optoacoustic imaging, and has the potential to improve pre-clinical and translational imaging applications. 
\end{abstract}


\section{Introduction}
\label{sec:introduction}
Optoacoustic (OA) imaging detects broadband ultrasound (pressure) waves generated within tissue in response to external illumination with light of transient energy, due to light absorption by tissue elements and thermo-elastic expansion. Using forward models that describe sound propagation in tissue, ultrasound measurements from multiple positions surrounding the object imaged are mathematically reconstructed to resolve the spatial distribution of the initial pressure rise. The reconstructed pressure rise is proportional to the product $H=\mu_a \phi$, whereby $\mu_a$ is the optical absorption coefficient and $\phi$ is the light fluence \cite{b1, b2, b2a}. The value {\it H} has only positive values in biological tissues since both absorption and light fluence are positive. However, the appearance of negative values is common in OA images due to different factors, such as the use of inaccurate forward models, inversion schemes, numerical errors, limited view detection geometry, transducer impulse response, unknown or unpredictable experimental effects or noise in the imaging system. The presence of negative values in the reconstruction does not have physical relevance. Importantly, when spectral techniques are employed, such as Multispectral Optoacoustic Tomography (MSOT) \cite{b3,b4}, the presence of negative values make spectral quantification problematic. 

It is therefore important to treat the appearance of negative values in the OA tomography problem. Model based reconstruction has been suggested as an alternative to back-projection algorithms to improve the accuracy of OA imaging, further incorporating transducer and laser characteristics into the inversion procedure \cite{b5, b6, b6a, b6b}. In principle, accurate inversion can reduce the image artifacts, but errors persist due to different experimental challenges including limited-angle signal collection, limited bandwidth detection, noise and other uncertainties, leading to incomplete data problems and results in the presence of erroneous negative values \cite{b5, b7, b8, b9}. Consequently, methods to directly treat the problem of negative values have been considered \cite{b4, b10, b11}. Ding et. al. \cite{b10} compared the utility of different minimization procedures using non-negative constraints, including steepest descent, conjugate gradient, and quasi-newton based inversion. Typical non-negative constraint schemes truncate the negative values within each step of the gradient iteration, forcing a result containing only positive or zero values. This practice however may bias the solution and generate inaccuracies in the reconstruction.  

An alternative approach to address the problem of negative values is to use image content for image correction. Image features such as the total energy (smoothness), contrast, total variation of an image can be generally employed as prior information to direct the inversion towards pre-determined outcomes, usually based on the assumptions about the nature of the image. For example, $\ell_2$- or $\ell_1$-norm minimization of the total variation of an image minimizes the edges of the reconstructed image. Using this notion, negative artifacts can then be eliminated by applying an explicit non-negativity constraint along with $\ell_2$-norm minimization \cite{b10, b12}. Another image metric that has been considered for eliminating negative values is the entropy of an image \cite{b13, b14}. Entropy is the measure of randomness in an image. Randomness of the image implies that information from each subpixel is assumed to be independent of each other and can statistically take any value irrespective of its neighboring subpixel. This becomes very useful in limited data situations; wherein the principle of maximum entropy tries to eliminate all uncertainties within each subpixel (among the different possible solutions) by imposing independent statistical structure on each pixel. Maximization of entropy (i.e. maximizing the term $-\bf{x}log(\bf{x})$; whereby $\bf{x}$ is the vectorized image) is equal to minimizing the term $\bf{x}log(\bf{x})$ and is a method considered in  Positron Emission Tomography (PET) and multi-modal imaging \cite{b13, b14} or astronomical imaging \cite{b15}. 
 
In this work, we examine the use of entropy as a prior in OA image inversion, in the context of nonlinear conjugate gradient minimization \cite{b16}. We hypothesize that the use of an entropy-based prior, which implements an implicit non-negativity constraint, can improve the accuracy of OA inversions over externally imposed non-negativity constraints. To prove this hypothesis, we first theoretically compare a conventional $\ell_2$-norm minimization problem using a smoothness constraint to an entropy maximization problem. We show that images reconstructed by entropy maximization cannot take negative values. The reconstructed OA images were further improved by correcting for the fluence, the fluence was estimated using finite volume method after segmenting the imaging domain (phantom or mouse). Thereafter, we compare the performance of inversion (after fluence correction) using entropy maximization and conventional inversion with externally applied non-negativity constraint using numerical simulation, experimental phantoms and small animal imaging. We discuss the performance differences observed and the advantages and limitations of using entropy maximization.

\section{Materials and Methods}

\subsection{Theoretical  background}
The propagation of the acoustic pressure wave generated due to the short-pulsed light absorption is governed by the following inhomogenous wave equation \cite{b17},
\begin{equation}
    \frac{\partial^2 p(r,t)}{\partial t^2} - c^2 \rho \nabla.(\frac{1}{\rho} \nabla p(r,t))= \Gamma \frac{\partial H(r,t)}{\partial t},
    \label{Eq:1}
\end{equation}
where the instantaneous light power absorption density in $\frac{W}{m^3}$ is indicated by $H$ and $\Gamma$ represents the medium-dependent dimensionless Grueneisen parameter. In Eq. \ref{Eq:1}, the tissue density is represented by $\rho$ while $c$ indicates the speed of sound (SoS). For our experiments, a uniform SoS of 1520 m/sec was heuristically estimated using image autofocusing method \cite{b18}. The initial pressure rise at position $r$ and time $t$ is given as $p(r,t)$. The solution for the wave equation can then be obtained using a Green's function by assuming $H(r,t)=H_r(r) \delta(t)$, which results in \cite{b17},
\begin{equation}
    p(r,t) = \frac{\Gamma}{4 \pi c} \frac{\partial}{\partial t} \int_{R=ct} \! \frac{H_r(r')}{R} \, \mathrm{d}r',
\label{Eq:2}
\end{equation}
where $R=ct$ represents the radius of the integration circle over a line element given as $dr'$. The above solution is subsequently discretized into the following matrix equation \cite{b19},
\begin{equation}
    b=\bf{A}x,
    \label{Eq:3}
\end{equation}
where $b$ is the boundary pressure measurements, $\bf{A}$ is the interpolated model matrix and $x$ is the unknown image to be reconstructed, representing the initial pressure rise distribution. The above formulation represents the forward model, i.e. given the initial pressure rise one can estimate the pressure at the boundary locations detected by the transducers. Thus, the acoustic inverse problem involves reconstructing the initial pressure rise given the boundary pressure data. In the $\ell_2$-norm formulation, the inverse problem is solved by minimizing a function given as, 
\begin{equation}
    \Omega_{\ell_2} = \operatornamewithlimits{arg\ min}\limits_{x} (||{\bf{A}}x-b||_2^2 + \lambda||{\bf L}x||_2^2),
    \label{Eq:4}
\end{equation}
where $\lambda$ is the regularization parameter. The term $||{\bf A}x-b||_2^2$ is called the residual term. The term $||{\bf L}x||_2^2$ is a $\ell_2$-norm of the second order total-variation of the image $x$ and ${\bf L}$ indicates the Laplacian operator. The value of the regularization parameter affects the resolution characteristics of the reconstructed image; higher the value of regularization the smoother the reconstructed image. 

\subsection{Entropy Maximization and Non-negative constraint}
An alternative method to the minimization problem of Eq. \ref{Eq:4} (optoacoustic reconstruction), is maximization of the entropy of the image. To elaborate on this point, a statistical approach is considered, wherein we assume that the image to be reconstructed follows a Gaussian distribution with estimated mean and standard deviation values. The dimension of the image to be reconstructed is $N \times N$, i.e. a vector of size $NN (=N^2)$. Next, we assume that each pixel $j$ in this image will be formed by a group of subpixels indicated by $m_j$ ($>=1$) and $M=\sum_{j=1}^{NN} m_j$. With these assumptions, let us consider the following experiment: wherein $K$ particles are distributed over all subpixels and let $K_i$ be the number of particles that fall in pixel $i$. Then the number of combinations to place $K$ particles in $NN$ pixels such that $K_j$ particles are present in pixel $j$ is given as,
\begin{equation}
    C(\tilde{K}) = \frac{K!}{\Pi_{j=1}^{NN} K_j!},
    \label{Eq:5}
\end{equation}
Further we have $m_j^{(K_j)}$ ways to put $K_j$ particles into $m_j$ subpixels. Hence, the total number of combinations to create the particle distribution $V(\tilde{K})$ is given as,
\begin{equation}
    V(\tilde{K}) = C(\tilde{K}). \Pi_{j=1}^{NN} m_j^{(K_j)},
    \label{Eq:6}
\end{equation}
The total number of particles in the distribution is given as $M^K$. Now making the assumption that each particle is equally likely i.e. uniform distribution. We get the probability of distribution of $\tilde{K}$ as,
\begin{equation}
    p(\tilde{K}) = \frac{V({\tilde{K}})}{M^K},
    \label{Eq:7}
\end{equation}
Now using Stirling approximation i.e. $K! \approx K^K e^{(-K)}$, we can write,
\begin{equation}
    log(p(\tilde{K})) = -K \sum_{j=1}^{NN} z_j log(\frac{z_j}{\hat{m}_j}),
    \label{Eq:8}
\end{equation}
where $z_j=\frac{K_j}{K}$ and $\hat{m_j} = \frac{m_j}{M}$. The average value inside a pixel $x_j$ will now be proportional to $z_j$ i.e. $x_j=S.z_j$ and $\tilde{m_j} = S.\hat{m_j}$ such that,
\begin{equation}
    \sum_{j=1}^{NN} x_j = \sum_{j=1}^{NN} \hat{m_j} = S,
    \label{Eq:9}
\end{equation}
with $x_j \ge 0$, $\hat{m_j}>0$.

Now let the prior distribution of the image vector be considered as $p_A (x)$, which is given as,
\begin{equation}
    log(p_A (x)) = -\frac{K}{S} \sum_{j=1}^{NN} x_j log(\frac{x_j}{m_j}),
    \label{Eq:10}
\end{equation}
which follows the relative entropy definition and is always non-negative (not defined for negative values). Our next assumption is that the error vector or the noise is normally distributed with zero mean and standard deviation $\sigma$ given as,
\begin{equation}
    p(r_i) = c.e^{\frac{-r_i^2}{2\sigma^2}},
    \label{Eq:11}
\end{equation}
which can be rewritten as,
\begin{equation}
    p(y|x) = c.e^{\frac{||{\bf A} x - b||_2^2}{2\sigma^2}},
    \label{Eq:12}
\end{equation}
Rewriting the overall expression using Bayes rule we get,
\begin{equation}
    log(p(x|y)) = -\frac{K}{S} \sum_{j=1}^{NN} x_j log(\frac{x_j}{m_j}) - \frac{1}{2\sigma^2}{||\bf{A} x - b||_2^2},
    \label{Eq:13}
\end{equation}
Neglecting the terms independent of $x$. We can pose this as an entropy maximization problem which is non-linear convex maximization problem, and this can be solved by minimizing the function,
\begin{equation}
    \Omega_{maxent} = \operatornamewithlimits{arg\ min}\limits_{x} (||{\bf{A}}x-b||_2^2 + \lambda \sum_{i=1}^{NN} x_i log(\frac{x_i}{m_i})),
    \label{Eq:14}
\end{equation}
where $-xlog(\frac{x}{m})$ indicates the relative entropy function of image $x$, typically $m$ is assumed to be an arbitrary constant \cite{b15}. In this work $m$ is assumed to be 1. Detailed mathematical analysis on the use of Eq. \ref{Eq:14} for applying an implicit non-negativity constraint, stability, and convergence of entropy maximization is given in \cite{b22}. Herein we study how positive values are retained with entropy maximization scheme. 

In $\ell_2$-norm minimization (Eq. \ref{Eq:4}), the gradient update equation at iteration $i$ is given as,
\begin{equation}
    x_i = x_{i-1} - ({\bf A}^T ({\bf A} x_{i-1} - b)) - \lambda {\bf L}^T {\bf L} x_{i-1},
    \label{Eq:15}
\end{equation}
The above update equation is obtained by taking the derivative of the objective function in Eq. \ref{Eq:4}. Note that in the above equation all the quantities will always be in real space i.e. ($\bf{A}$, $x_{i-1}$, $x_i$, $b \ \in \ {\rm I\!R} $), and can take any values due to the absence of any natural non-negativity barrier. Therefore, the $\ell_2$-norm based minimization can generate negative values (which can be in ${\rm I\!R}$) during the image reconstruction procedure. In case of entropy maximization (Eq. \ref{Eq:14}), the gradient updated equation at iteration $i$ is given as,
\begin{equation}
    x_i = x_{i-1} - ({\bf A}^T ({\bf A} x_{i-1} - b)) - \lambda (1 + log(\frac{x_{i-1}}{m_{j-1}})),
    \label{Eq:16}
\end{equation}
The derivation pertaining to applying implicit positivity constraint using entropy maximization is discussed in the Appendix-I.

Choice of regularization plays a key role in reconstructed image quality by defining over-smoothed or under-smoothed approximations in case of $\ell_2$-norm based reconstruction. In terms of distance measure, $\ell_2$-norm constraint can be considered as Euclidean distance between the prior and the expected image, i.e. $||{\bf L}x||_2 = <{\bf L}x,{\bf L}x> \Longleftrightarrow <{\bf L}x,{\bf L}x_{pr}>$\cite{b23,b24,b25}, therefore higher regularization will weigh the $\ell_2$-norm constraint more and thus resulting in a smoother solution. Similarly entropy maximization can be related to Kullback-Leiber distance, as cross entropy between prior and the expected image, i.e. $\sum x log(x) = \sum x log(\frac{x}{x_{pr}})$, therefore higher regularization will push the subpixels (i.e. $x_{pr}$) in pixel $i$ of image vector $x$ to uniform distribution \cite{b23,b24,b25}. Thus, low regularization in the entropy maximization scheme will result in minimizing the residual (i.e. noisy reconstruction), whereas choosing higher regularization will result in the initial pressure rise being close to a smooth distribution having intrinsically positive values. The operating range of the regularization parameter in the entropy maximization framework can be found using the L-curve type method, cross-validation based scheme \cite{b26,b27}.

\subsection{Choice of regularization parameter - L-curve method}
Typically, the regularization parameter ($\lambda$) is chosen automatically using the L-curve method \cite{b20,b21}. 
The L-curve method is a popular method for automatically choosing the regularization parameter for a linear inverse problem and this scheme was earlier used in diffuse optical tomography and OA tomography. In the L-curve method, a graph is plotted between the residual ($||{\bf{A}}x-b||_2^2$) and the reconstruction ($||x_{\lambda}||_2^2$) as function of regularization parameter ($\lambda$). This essentially means that the reconstructed solution ($x_{\lambda}$) is a function of regularization ($\lambda$). In an ideal case this curve will be of L-shape. The corner point of this L-shape represents the least distance from the origin, indicating an ideal balance between residual and expected solution.
For the case of entropy maximization the solution norm will be replaced by entropy term i.e. ($\Sigma {x_{\lambda} log(x_{\lambda}})$).
In this work, we use L-curve type approach to automatically estimate the regularization parameter in both the L2-norm and entropy maximization schemes. 

\subsection{$\ell_2$-norm with smoothness and non-negativity constraint}
Minimizing the function in Eq. \ref{Eq:4} was performed using a conjugate gradient method (equivalent to iterative least squares QR (LSQR) method), which has a closed form solution as \cite{b28},
\begin{equation}
    x \approx x_{\ell_2-lsqr} = \bf{V_k} (\bf{B_k}^T \bf{B_k} + \lambda \bf{S_k}^T \bf{S_k})^{-1} \beta_0 \bf{B_k}^T e_1,
    \label{Eq:17}
\end{equation}
where $\bf{B_k}, \bf{S_k}, \bf{V_k}, \beta_0$, and $e_1$ can be obtained in the Lanczos diagonalization procedure with $\begin{pmatrix}
   {\bf A} \\\
\lambda {\bf L}
 \end{pmatrix}$ and $\begin{pmatrix}
   b \\\
 0
 \end{pmatrix}$. Here $k$ indicates the number of iterations during the joint bidiagonalization procedure. 

In the $\ell_2$-norm formulation with non-negativity constraint, the following minimization is solved, 
\begin{equation}
    \Omega_{\ell_2-NN} = \operatornamewithlimits{arg\ min}\limits_{x} (||{\bf{A}}x-b||_2^2 + \lambda||{\bf L}x||_2^2) \ \ \ \ \ \ \ \ \bf{s.t.}\ \ \ x>0,
    \label{Eq:18}
\end{equation}
The above minimization is solved using the LSQR solver and then the obtained solution containing negative values are thresholded to 0, as negative values do not have any physical relevance (as optical absorption coefficient in biological tissue is not negative). Eq. \ref{Eq:17} is used to obtain the solution and then the negative values in the solution are thresholded. The regularization parameter was chosen using L-curve method (explained in Sec. II-C)\cite{b20}.

\subsection{Implementation Steps for Entropy Maximization}
Eq. \ref{Eq:14} is minimized using a non-linear conjugate gradient type method and the step-length for the conjugate gradient method is computed using a line search \cite{b29}. Minimization of the objective function in Eq. \ref{Eq:14} with conjugate gradient requires computing the derivative and then move in independent perpendicular gradient direction. The derivative used in the conjugate gradient scheme for the objective function in Eq. \ref{Eq:14} is computed as,
\begin{equation}
    \nabla \Omega_{maxent} = 2{\bf A}^T (\bf{A} x -b) + \lambda (1 + log(\frac{x_{i-1}}{m_{i-1}})),
    \label{Eq:19}
\end{equation}
The minimization is presented in more details in the Algorithm-1 section. The regularization parameter was chosen using an L-curve method (as a tradeoff between negative of entropy and residual).

\begin{algorithm}
\caption{\emph{Entropy Maximization Algorithm} }
  AIM: Estimation of $x$ in Eq. \ref{Eq:14}\\
  INPUT: Obtained boundary pressure data ($b$), Interpolated Model Matrix (${\bf A}$), Regularization Parameter ($\lambda$), Initial Guess ($x_0$).\\
 OUTPUT: Reconstructed Initial Pressure Rise ($x$) \\
 Initialize: Iteration Number ($iter=0$), Tolerance ($tol=1e^{-8}$), $\omega = 0.5$, Maximum Iterations ($max_{iter}$) = 500
 \hrule

   1. Compute Gradient ($g(x) = 2 {\bf A}^T ({\bf A} x_0 - b) + \lambda (1 + log(x_0)$), Residue ($r = {\bf A} x_0 - b$), $p = -g$, $\Phi_0 = p^T g$, $x_{prev} = x_0$, $g_{prev} = g$, $\Delta x = x_{prev}$\\
while $iter<max_{iter} \ \& \ \Delta x < (tol \times ||x||_2)$ 
\begin{itemize}
\item[1.] ${A_p} = {\bf A} \times p$, $\gamma = A_p^T A_p$, $v={\bf A}^TA_p$, $t = 1, u = 1$
\item[2.] Improve step-length ($\alpha$) to ensure descent direction traversal; while  $u > -\omega \times t$
\begin{itemize}
\item[1.] $\Phi = \Phi_0 + 2 \alpha \gamma + \lambda p^T (1 + log(\frac{\alpha p}{x_{prev}}))$; ($\alpha$ is estimated using secant root finding method such that $\Phi(\alpha) = 0$).
\item[2.] Update gradient: $g_{temp} = g_{prev} + \lambda(1+log(\frac{\alpha p}{x_{prev}})) + 2\alpha v$
\item[3.] Update CG variables: $\beta = \frac{g_{temp}^T g_{temp} - g_{prev}^T g_{prev}}{\Phi-\Phi_0}$, $t = g_{temp}^T g_{temp}$
\item[4.] $u = -g_{temp}^T g_{temp} + \beta \Phi$
\end{itemize}
end
\item[3.] Update the solution: $g_{prev} = g_{temp}$, $\Delta x = \alpha p$, $x_{prev} = x_{prev} + \Delta x$
\item[4.] Update cost, residue and gradient information: $p = -g_{prev} + \beta p$, $r = r + \alpha A_p$, $\Phi_0 = p^T g$
\end{itemize}
end \\
2. Final solution: $x = x_{prev}$
 \end{algorithm}

\subsection{Fluence Correction}
The image reconstructed in Eq. \ref{Eq:17} (LSQR) and with Algorithm-1 (Entropy Maximization) represents the absorbed energy distribution $H_r (r)$ in tissue, which depends on the fluence distribution and the optical absorption coefficient  $\mu_a (r)$ i.e \cite{b17},
\begin{equation}
    x = p_0 (r) = H_r (r) = \mu_a (r) \Phi (r),
    \label{Eq:20}
\end{equation}
where $p_0 (r)$ is the initial pressure rise distribution and $\Phi(r)$ indicates the local light fluence density in $mJ/cm^2$. To extract the absorption coefficient map, it is therefore critical to estimate the fluence in the medium imaged. Different schemes have been developed for estimating the fluence distribution and quantitatively recover optical absorption coefficient maps, including model-based inversion schemes integrated with fluence compensation \cite{b30a}, wavelet frameworks \cite{b30b}, finite-element implementation of the delta-Eddington approximation to the radiative transfer equation \cite{b30c}, diffusion equation based regularized Newton method \cite{b30d}, or approximations with base spectra \cite{b31}. Herein we assumed for demonstration purposes a light propagation model based on the diffusion equation, further assuming that scattering dominates over absorption \cite{b32}, which is a valid approximation for  most biological tissues and NIR measurements, i.e.,
\begin{equation}
    -\nabla.[D(r).\nabla \Phi(r)] + \mu_a(r) \Phi(r) = S_0 (r),
    \label{Eq:21}
\end{equation}
where $D(r)=\frac{1}{(3(\mu_a+\mu_s^{'}))}$ is the diffusion coefficient and $\mu_s^{'}(r)$ indicates the reduced scattering coefficient at position $r$. $S_0 (r)$ indicates the light source at the boundary of the imaging domain. Eq. \ref{Eq:21} is used for fluence estimation, and the diffusion equation is solved using the finite volume method (FVM). Optical properties were based on the known phantom specifications or estimates of absorption and scattering coefficients of tissue from the literature \cite{b33}. Then, we obtained absorption coefficient maps by normalizing the images with the corresponding calculated fluence distribution \cite{b34}. Since OA measurements of phantoms were performed in a water bath, we also employed the Beer-Lambert Law ($OD = -log(\frac{I}{I_0}) = -\mu_a d$) to model photon propagation in water. The relative distances in phantom and water were assigned after segmentation of the OA images. The entire workflow of segmentation and fluence correction is integrated with the proposed non-negative entropy maximization algorithm to render improved image quality.

\subsection{Imaging instrumentation and protocol(s):}
Experimental data was acquired using the multispectral optoacoustic tomography (MSOT) scanner \cite{b35} (MSOT256-TF, iThera Medical GmbH, Munich, Germany). The boundary pressure readouts (time-series) were collected at 2,030 discrete time points at 40 Mega samples per second using a 256-element cylindrically focused transducer, resulting in the number of measurements ($M$) being 2030x256=519,680. The utilized piezocomposite transducer had a central frequency of 5 MHz with a radius of curvature of about 40 mm and an angular coverage of 270$^{\circ}$. Uniform illumination was achieved with a ring type of light delivery using laser fiber bundles. Numerical simulations were performed with the same configuration as MSOT256-TF system with a realistic breast phantom having spatially varying absorption coefficient (in $cm^{-1}$) as shown in Fig. 1(a). Next, we segmented the boundary of the breast region in Fig. 1(a) and estimated the fluence distribution (shown in Fig. 1(b)) by solving the hybrid model (Sec. II.F) with the absorption coefficient and reduced scattering coefficient set to 0.2 $cm^{-1}$ and 12 $cm^{-1}$ respectively. The initial pressure rise (in $kPa$) was then estimated by multiplying the fluence distribution (Fig. 1(b)) with the spatially varying optical absorption (Fig. 1(a)), the initial pressure rise distribution (after scaling with acoustic parameters) is shown in Fig. 1(c). Note that we assumed point detector and did not model transducer characteristics in the simulations. The numerical breast phantom was created by using contrast enhanced magnetic resonance imaging \cite{b36}. Eq. 3 was used to model the acoustic propagation (on a $512\times512$ grid) and the pressure signals were collected at specific detector locations. The model matrix in Eq. 3 was built using interpolated model matrix method as explained in  Ref. \cite{b19}. To avoid inverse crime, the simulated data was generated on an imaging grid of size 512x512, while the reconstruction was performed on a grid of size 256x256. The simulated data was added with additive white Gaussian noise, to result in a SNR of 32 dB in the simulated data.

To verify the quantitative reconstruction capabilities of the proposed entropy maximization scheme, a star shaped (irregular) phantom was created. The phantom constituted of a tissue mimicking (7\% by volume of Intralipid and pre-computed volume of diluted India ink added) agar core having the optical density of 0.25. Two tubular absorbers made up of India-ink with the absorption coefficient values of 2.5 OD (calibrations done with Ocean Optics USB 4000) were inserted in the phantom. The absorbers were placed at two different depths within the phantom (one at the center and the other at the edge of the imaging domain) to test the sensitivity of the proposed scheme in reconstructing the absorbers at different imaging distances from the sensing arrays. Under normal operating conditions, the fluence at the center of the imaging domain is significantly lower as compared to the boundary of the object imaged, owing to the optical attenuation of the incident irradiation. Hence, performing fluence correction becomes indispensable to assign appropriated intensity to the absorber at the center of the imaging domain. 

The proposed methods were further validated on {\it in-vivo} mouse abdomen and brain datasets drawn from a standardized {\it in-vivo} murine whole body imaging database (10 mice/30 anatomical datasets) previously developed in Ref. \cite{b18}. The selected images were obtained at a laser wavelength of 760 nm and 800 nm, and the water (coupling medium) temperature was maintained at 34$^{\circ} C$ for all experiments. Non-negativity based entropy maximization scheme was further validated using spectral measurements. Spectral measurements were acquired from a tumor bearing nude BALB-C mice with the laser wavelengths running from 680 nm to 900 nm at steps of 20 nm. All animal experiments were conducted under supervision of trained technician in accordance with institutional guidelines, and with approval from the Government of Upper Bavaria.

\subsection{Figure of merit}
To develop an objective approach to evaluate imaging performance of different reconstruction methods, we used line plots on the reconstructed image (from phantom and tissue measurements). We also performed quantification using sharpness metric, defined as,
\begin{equation}
    SM = \frac{\sum \frac{dI^2}{dx^2} + \frac{dI^2}{dy^2}}{n},
    \label{Eq:22}
\end{equation}
The sharpness metric indicates the edges in the reconstructed image ($I$): the higher the value of $SM$, the sharper the reconstructed image. This figure of metric was used for evaluating the proposed method, as the non-negative constraint tend to introduce zeros in the reconstructed image. The number of non-negative values is also reported for comparing the different reconstruction methods.  Note that the number of negative pixels were calculated from the phantom or mice region (excluding the water region).

Further root mean square error (RMSE) and peak signal to noise ratio (PSNR) was used to evaluate the performance of different reconstruction methods with numerical simulation. RMSE is given as,
\begin{equation}
    RMSE = \sqrt{\frac{\sum_o (x_o^{recon} - x_o^{true})^2}{NN}},
    \label{Eq:23}
\end{equation}
is computed for comparing the performance of different algorithm. Here $x_o^{true}$ is the $o^{th}$ pixel of ground truth and $x_o^{recon}$ is the $o^{th}$ pixel of reconstructed image. PSNR is defined as,
\begin{equation}
    PSNR = 20 \times log(\frac{max(x^{true})}{RMSE}),
    \label{Eq:24}
\end{equation}

The calculated sharpness metrices (for phantom and \textit{in vivo} small animal images), and the $RMSE/ PSNR$ values (for simulations) are given in section \ref{sec:results}. 

\section{Results}
\label{sec:results}

Fig. 1(c) shows the initial pressure distribution with the realistic numerical breast phantom used to evaluate the performance of different reconstruction methods. The reconstructed initial pressure rise distribution using the $\ell_2$-norm based reconstruction is shown in Fig. 1(d). The solution pertaining to $\ell_2$-norm based reconstruction (along with non-negative constraint) is indicated in Fig. 1(e). The reconstructed optoacoustic image using the entropy maximization approach is represented in Fig. 1(f). The reconstructions containing negative values are indicated with a red colormap, hence the negative pixels in Fig 1(d) are shown in red color. From the numerical simulations, it is apparent that the $\ell_2$-norm based reconstruction produces negative values by just adding noise to the data and incorporating fluence effects, however these negative values do not appear after thresholding and using entropy maximization scheme as indicated by red arrows in Figs 1(e) and 1(f). Furthermore $\ell_2$-norm with thresholding results in a nosier reconstruction with limited structures compared to entropy maximization scheme as shown with red arrows in Figs 1(e) and 1(f). The PSNR values for $\ell_2$-norm, $\ell_2$-norm with thresholding and entropy maximization reconstruction are 29.9736 dB, 30.2616 dB and 30.3529 dB respectively. The RMSE values for $\ell_2$-norm, $\ell_2$-norm with thresholding and entropy maximization reconstructions are 0.0453, 0.0451, and 0.0450 respectively. The number of reconstructed negative pixels with $\ell_2$-norm reconstruction with numerical breast phantom is 4370. Note that the simulation studies did not model many experimental parameters like impulse response of the transducer, physical dimension of the transducer, pitch of the detector, artifacts arising due to reflections, and these parameters are known to influence the OA measurements in experimental scenarios. Further, we proceeded to study the performance of the proposed entropy maximization scheme with phantom and \textit{in-vivo} datasets.
\begin{figure}[!h]
\centerline{\includegraphics[width=\columnwidth]{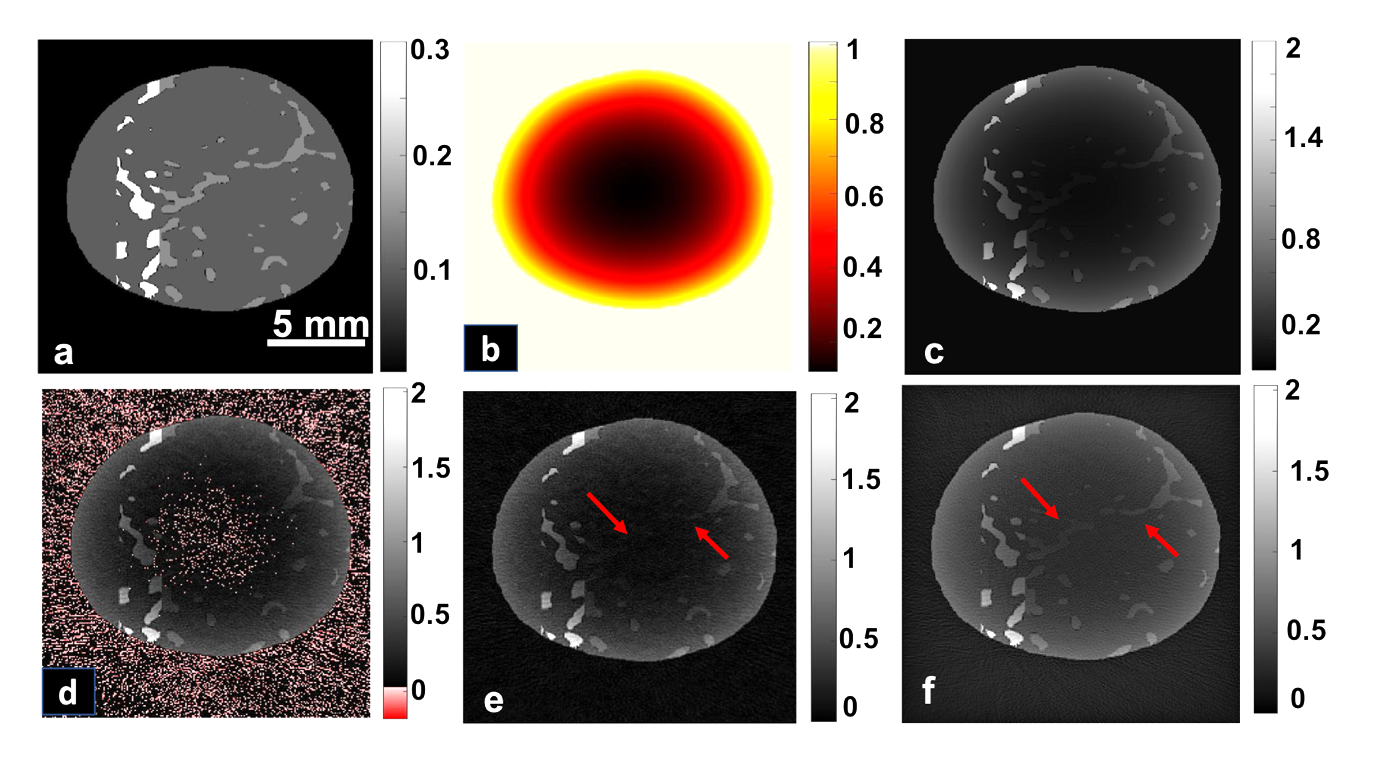}}
\caption{Comparative evaluation of entropy maximization scheme with standard non-negative reconstruction using numerical simulations.  (a) shows the absorption distribution of the used numerical breast phantom, (b) shows the fluence distribution in the imaging domain, and (c) indicates the initial pressure rise distribution of the numerical breast phantom. Reconstructed initial pressure rise image of numerical breast phantom using the (d) $\ell_2$-norm based reconstruction, (e) $\ell_2$-norm based reconstruction with thresholding, (f) entropy maximization reconstruction. The negative values are plotted in a different colormap (d) for visualization and colormaps indicate quantitative values.}
\label{fig1}
\end{figure}

Fig. 2 shows reconstructions of the star phantom, which reveal the efficacy of the proposed method vis-a-vis traditional $\ell_2$-norm based reconstruction in generating positive values for both the initial pressure rise and absorption coefficient distribution. The reconstructed initial pressure rise and absorption coefficient distribution using the $\ell_2$-norm based reconstruction is shown in Figs 2(a) and 2(d) respectively. The reconstructed initial pressure rise and absorption coefficient distribution using the $\ell_2$-norm based reconstruction (with non-negative constraint) is indicated in Figs 2(b) and 2(e) respectively. The reconstructed initial pressure rise and absorption coefficient distribution using the entropy maximization based approach is represented in Figs 2(c) and 2(f) respectively. The reconstructions containing negative values are indicated with a red colormap, hence the negative pixels in Figs 2(a) and 2(d) are shown in red color.
\begin{figure}[!htb]
\centerline{\includegraphics[width=\columnwidth]{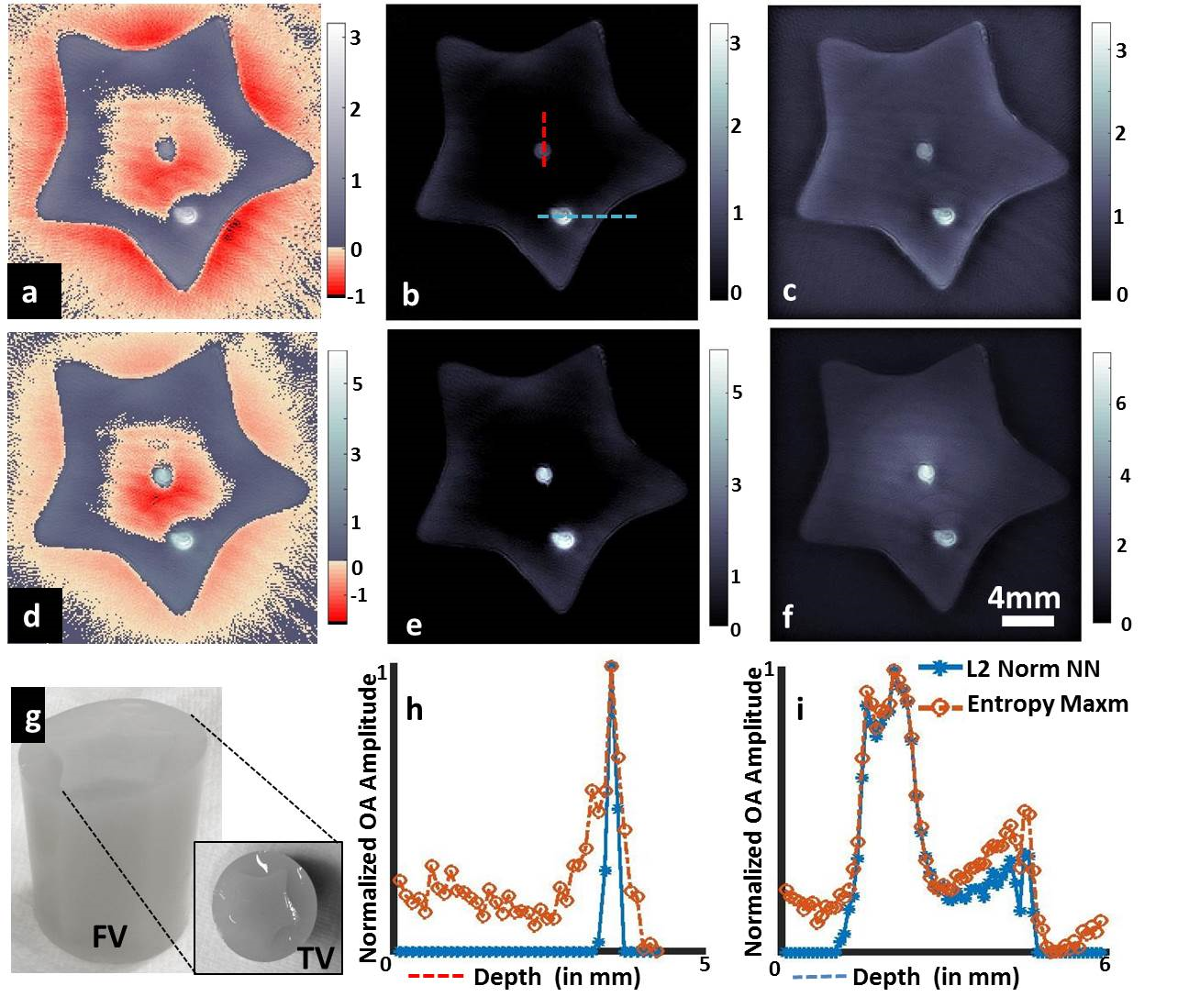}}
\caption{Comparative evaluation of entropy maximization scheme with standard non-negative reconstruction using phantom data.  Reconstructed OA image of star phantom using the (a) $\ell_2$-norm based reconstruction, (b) $\ell_2$-norm based reconstruction with thresholding, (c) entropy maximization based reconstruction. Absorption coefficient distribution after fluence correction using (d) $\ell_2$-norm based reconstruction, (e) $\ell_2$-norm based reconstruction with thresholding, (f) entropy maximization based reconstruction. (g) shows the photograph of the phantom used, (h) line profile along the vertical red dashed line indicated in 2(b), (i) line profile along the horizontal blue dashed line indicated in 2(b). The negative values are plotted in a different colormap (a and d) for visualization and colormaps indicate quantitative values (in a.u). }
\label{fig2}
\end{figure}

The proposed entropy maximization method (Fig. 2(f)) can provide accurate image representation with the ability to reconstruct the absorber (having OD of 2.5) at the center and the edge of the imaging domain along with reconstructing a star shaped background (having OD of 0.25). The negative values obtained using LSQR inversion is shown as red color in Fig. 2(a) and Fig. 2(d). The non-negative based $\ell_2$-norm reconstruction is able to generate reconstruction results with positive values, but is not able to correctly reconstruct the internal volume of the star (tissue mimicking agar with 0.25 OD) phantom which is accurately reconstructed using entropy maximization. Fig. 2(g) shows the photograph of the phantom used from front-view (FV) and top-view (TV). Fig. 2(h) indicates the line plot along the vertical red dashed line shown in Fig. 2(b). Fig. 2(i) indicates the line plot along the horizontal blue dashed line shown in Fig. 2(b). The sharpness metric and the number of non-negative values are shown in Table-\ref{tab1}. The quantitative metric indicate that the proposed method can provide accurate image representation. Fig. 2(f) and the line plots in Figs 2(h) and 2(i) demonstrate that the maximum entropy based scheme can deliver better contrast while maintaining the background intensity than the standard $\ell_2$-norm based reconstructions.The fluence correction was performed by using segmented (boundary) priors obtained automatically using deformable active contour models \cite{b37}. The results were corroborated with additional phantom (Agar block with 5\% intralipid) scans which included India ink insertions of 3 different ODs in tissue relevant concentrations - 0.15, 0.30 and 0.45 OD at 800nm measured using a spectrometer (VIS-NIR; Ocean Optics). The results demonstrate that the signal intensities change proportionately with the changing OD of the insertions, and the values are in agreement with other commonly used inversion algorithm (i.e Tikhonov). The reported signal intensities were obtained by taking the mean of the different ROI's indicated in Table-I of the supplementary. Additionally, the proposed reconstruction scheme recovered higher (absolute) signal intensities while reducing negative values in reconstructed image (see supplementary Table I).

Empirically selecting the regularization biases the reconstruction results.  Therefore an L-curve method was used to automatically choose the regularization parameter for Tikhonov method \cite{b20} and entropy maximization based scheme. Previous works have used L-curve approach for automatically choosing  the regularization parameter in entropy maximization framework for estimating distance distributions of magnetic spin-pairs \cite{b27}. Fig. 3 indicates the L-curve criterion used to choose the regularization parameter (details regarding L-curve approach is given in Sec. II-C) as applied to star phantom OA data presented in Fig. 2. Similar approach was used for automatically selecting the regularization parameter with numerical simulations and {\it in-vivo} data. Other methods like cross-validation can also be used for automatically choosing the regularization parameter in Tikhonov and entropy based framework \cite{b21, b26}. Further, we studied the effect of regularization parameter choice on reconstruction image quality. Fig. S1 in supplementary shows maximum entropy reconstruction at different regularization parameter values. It can be seen that at high regularization values, the solution leads to uniform distribution, however maximum entropy scheme seems to have a large operating range from 1 to 10,000.

\begin{figure}[!tb]
\centerline{\includegraphics[width=\columnwidth]{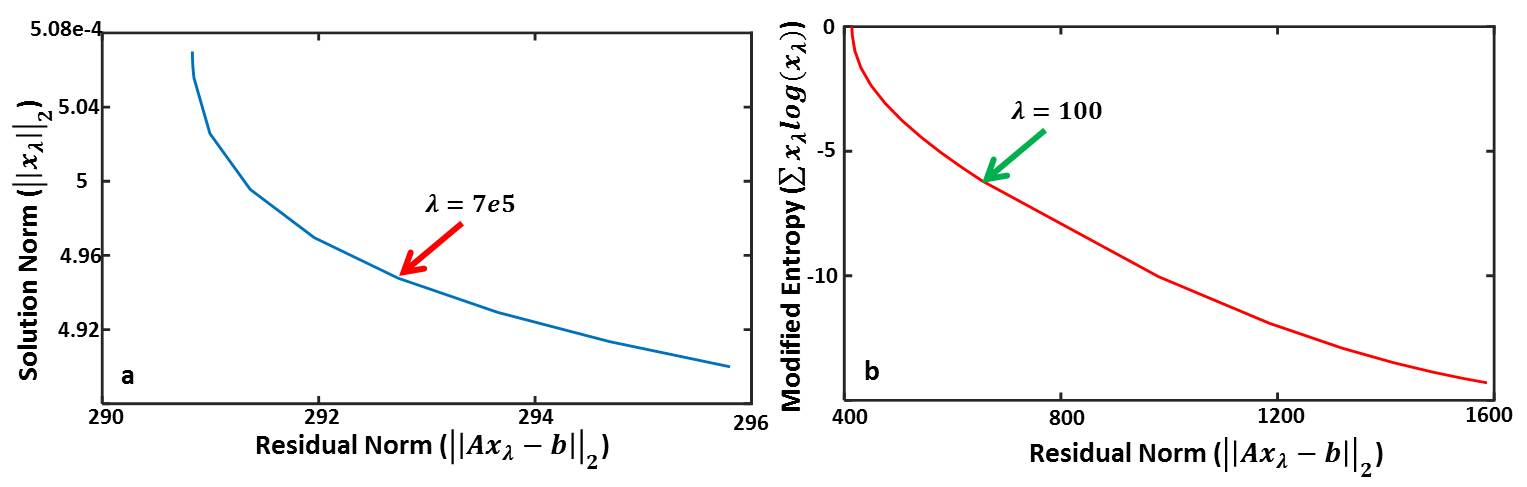}}
\caption{L-curve method for automatically choosing the regularization parameter (a) L-curve method for choosing the regularization parameter for Tikhonov based reconstruction (b) L-curve type approach for choosing the regularization parameter in the proposed entropy maximization scheme.}
\label{fig3}
\end{figure}

The maximum entropy based scheme depends on the initial guess used in the non-linear conjugate gradient scheme. The maximum entropy constraint involves a non-linear logarithmic term, and the logarithm of a negative value is not defined, therefore having a large positive value at the initial guess will always generates positive reconstruction distributions and thus plays an important role in intrinsically obtaining non-negative reconstruction. The same is elaborated in the Appendix-I. The reconstruction results corresponding to a backprojection-type initial guess ($\bf{A}^T b$ containing negative values; $\bf{A}^T$ indicates transpose of system matrix) is indicated in Fig. 4(a), the image shows the real part of the solution. The reconstruction results corresponding to the initial guess $(\frac{||b||_2}{||{\bf A}||_1} \times ones(NN,1))$ is indicated in Fig. 4(b). Fig. 4(a) clearly indicates that the negative values in the entropy maximization reconstructions arises because of initial guess used in the non-linear conjugate gradient scheme i.e. $(\frac{||b||_2}{||{\bf A}||_1} \times ones(NN,1))$ gives non-negative results while $\bf{A}^T b$ results in negative values.  Hence, in all the reconstructions the initial guess was chosen to be $(\frac{||b||_2}{||{\bf A}||_1} \times ones(NN,1))$ and the regularization parameter was chosen using the L-curve method. Note that reconstructions in Fig. 4 involve performing additional fluence correction. The colormap in the case of mouse images are normalized to maximum and minimum values and the negative values are indicated in red color. 
\begin{figure}[!htb]
\centerline{\includegraphics[width=\columnwidth]{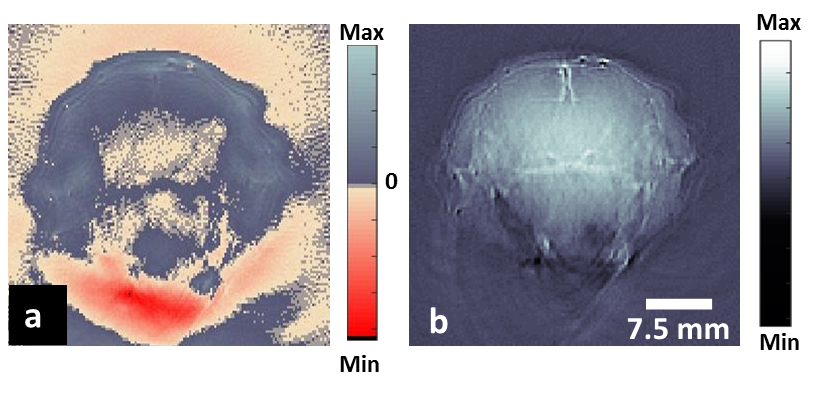}}
\caption{Dependence of initial guess on positivity constraint with entropy maximization scheme. Reconstructed optoacoustic image of mouse brain (head scanned {\it in-vivo}) using two different initial guesses in entropy maximization algorithm (a) $\bf{A}^T b$ (-ve values exists at initial guess) generates negative values and (b) $\frac{||b||_2}{||{\bf A}||_1} \times ones(NN,1)$ (only +ve value exist at initial guess) yields non-negative image. The negative values are plotted in a different colormap in (a) for visualization, colorbars indicates the absorption coefficient (in a.u). }
\label{fig4}
\end{figure}

Non-negative reconstruction generated with entropy maximization approach was further improved using fluence correction method. Fig. 5(a) shows the performance of segmentation approach in delineating the interface/boundary between the mice body (at the abdominal region) and water. The segmented boundary is used as a source term (after attenuation compensation using Beer-Lambert law in water) for modeling light propagation by solving the diffusion equation. Indeed, this boundary can be a good approximation for source term, as fiber bundle in the MSOT machine are arranged to provide uniform illumination on the sample. The fluence profile obtained after solving diffusion equation is shown in Fig. 5(b), the fluence was estimated with optical properties obtained from the literature \cite{b33}. Fig. 5(c) represents the initial pressure rise distribution reconstructed with entropy maximization approach. Fig. 5(d) shows the absorption coefficient distribution after normalizing the initial pressure distribution (Fig. 5(c)) with the estimated fluence profile (Fig. 5(b)). It can be clearly seen that signals from deeper regions on the mice gets highlighted more, similar approach was used for other regions of the mice. 
\begin{figure}[!htb]
\centerline{\includegraphics[width=\columnwidth]{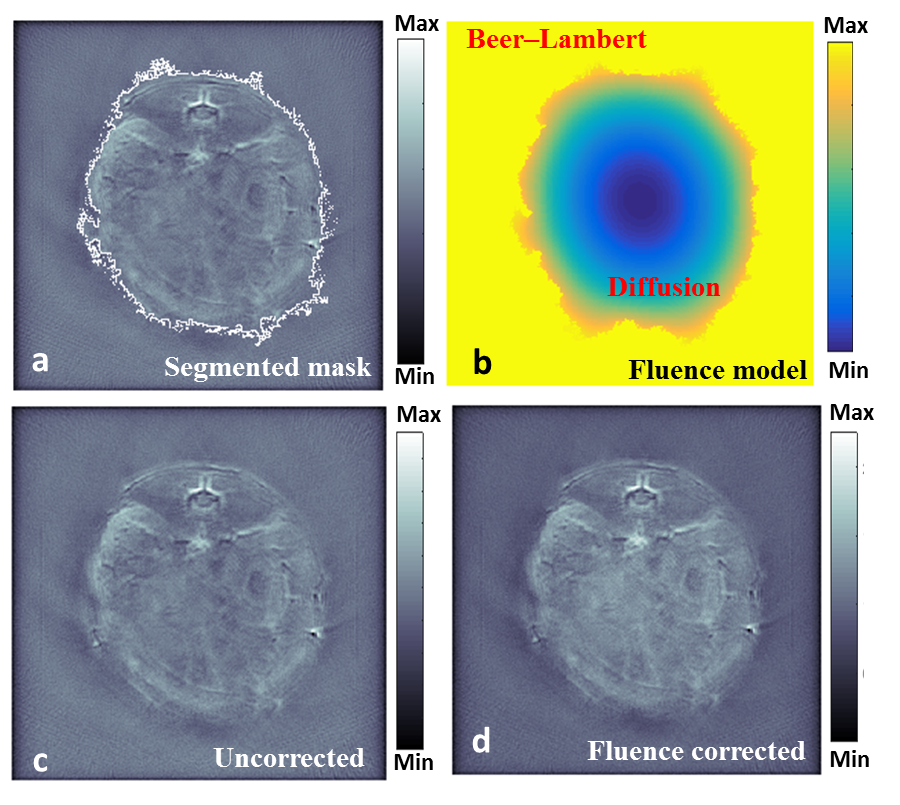}}
\caption{Improved optoacoustic reconstruction with segmented priors based fluence correction: (a) Segmentation mask estimated using active contours method for separating water and mouse (b) Fluence profile inside the mice region (c) Initial pressure distribution (in a.u.) reconstructed using the entropy maximization approach (d) Absorption coefficient distribution (in a.u.) after normalizing the initial pressure distribution (5(c)) with fluence profile (5(b)).}
\label{fig5}
\end{figure}

\begin{figure*}[!htb]
\centerline{\includegraphics[width=6in]{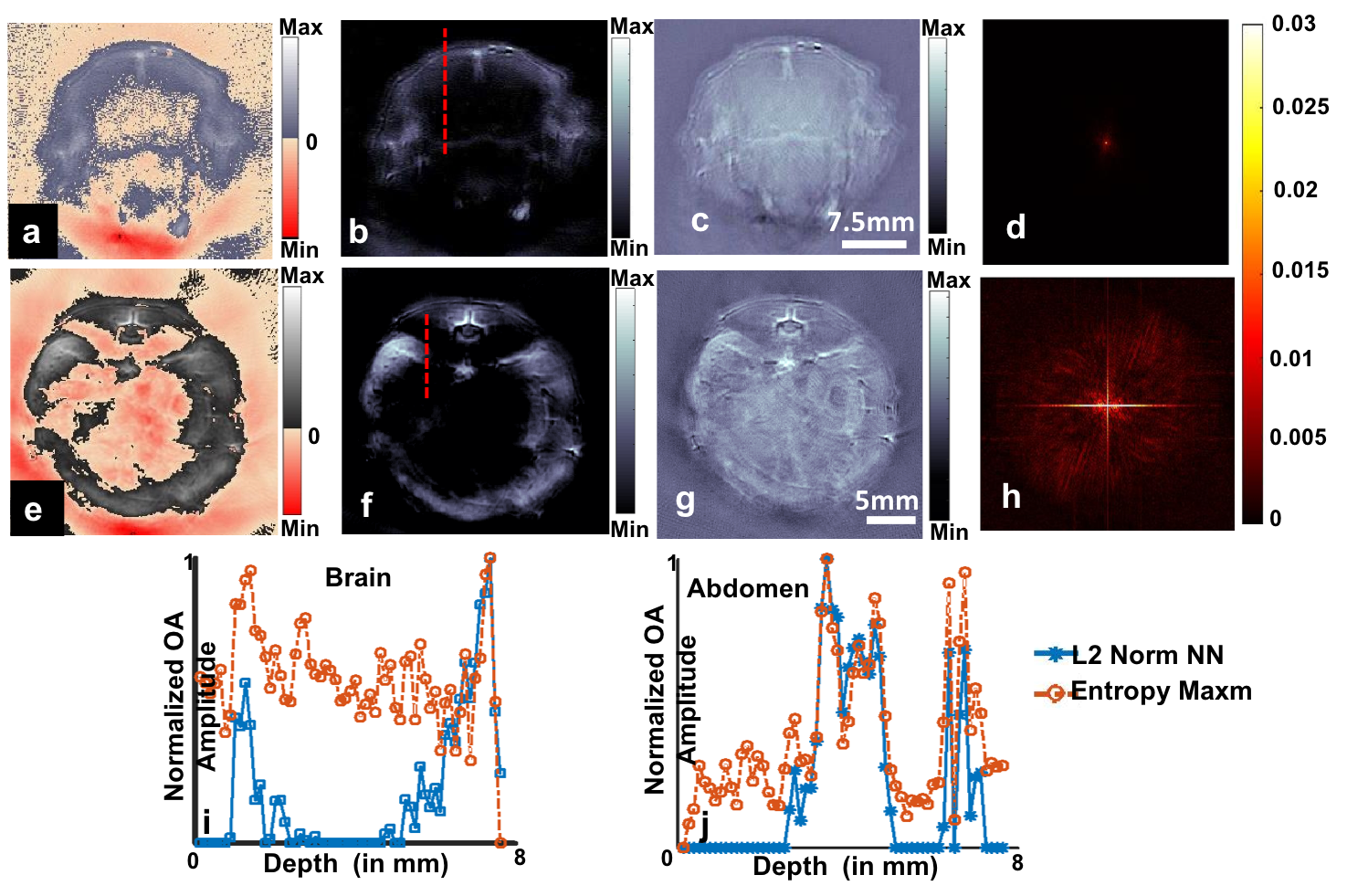}}
\caption{Comparison of entropy maximization scheme with standard non-negative reconstruction at two different mice regions. Reconstructed optoacoustic images using the (a) $\ell_2$-norm based reconstruction, (b) $\ell_2$-norm based reconstruction with thresholding, (c) entropy based reconstruction and fluence correction (using segmented prior) of murine head region; (d) represents the magnitude of Fourier domain signal for 6(f); Reconstructed optoacoustic images using the (e) $\ell_2$-norm based reconstruction, (f) $\ell_2$-norm based reconstruction with thresholding, (g) entropy based reconstruction (using segmented prior) for the mouse abdominal region imaged {\it in-vivo}. (h) represents the magnitude of Fourier domain signal for 6(g); (i) line profile along the red dashed line indicated in 6(b). (j) line profile along the red dashed line indicated in 6(f). The negative values appearing in $\ell_2$-norm based reconstruction scheme (a and e) are plotted in a different colormap (negative values marked in red) for visualization, colorbars indicates the initial pressure rise (in a.u). An 8 week old nude mice (CD-1® Nude, Charles River Laboratories, Germany) was imaged at an wavelength of 760 nm (brain) and 800 nm (abdomen). The negative values (if present in the reconstructed image) is marked with a different colormap.}
\label{fig6}
\end{figure*}

The reconstruction results (corresponding to absorption coefficient distribution) pertaining to the mouse head and mouse abdominal regions using the standard and proposed method are shown in Fig. 6. The reconstruction results corresponding to $\ell_2$-norm based scheme (solved using LSQR method) for the mouse head and abdominal region is indicated in Figs 6(a) and 6(e) respectively, and the corresponding results for $\ell_2$-norm based non-negative scheme (solved using LSQR method with thresholding) are given by Figs 6(b) and 6(f) respectively. The reconstruction results using the entropy maximization approach (Algorithm-1 with the integrated hybrid fluence correction) for the same anatomical regions is shown in Fig. 6(c) and Fig. 6(g) respectively. The experimental phantom and in-vivo reconstructions were performed on a 200x200 pixel imaging domain which corresponds to a physical field of view of 20mm x 20mm. The optical properties used for fluence estimation was assumed to be homogenous inside the tissue and taken from literature  \cite{b33}. Figs 6(d) and 6(h) indicate the Fourier domain representation of the reconstructed images (i.e. Fig. 6(f) and 6(g)) using L2-norm with thresholding and entropy maximization schemes respectively. We could clearly see that entropy maximization scheme (Fig. 6(h)) has more low frequency content when compared to L2-norm with thersholding (Fig. 6(d)). Fig. 6(i) indicates the line plot along the red dashed line shown in Fig. 6(b) and Fig. 6(j) shows the line plot along the red dashed line indicated in Fig. 6(f). The sharpness metric and the number of non-negative values for these reconstructions are indicated in Table-\ref{tab1}. These metrics show that the proposed method can provide accurate image reconstruction with lesser negative values and increased sharpness. Negative values should not arise during standard OA data acquisition, hence the lesser the number of negative pixels more accurate is the reconstructions. However in some scenarios the presence of negative values might indicate accurate reconstruction like temperature dependent studies \cite{b38}. However, we are working with standard OA acquisition, and thus more positive values indicate accurate reconstruction. Again, the colormap is normalized to maximum and minimum values, while indicating the negative values in red color.

\begin{figure}[!htb]
\centerline{\includegraphics[width=\columnwidth]{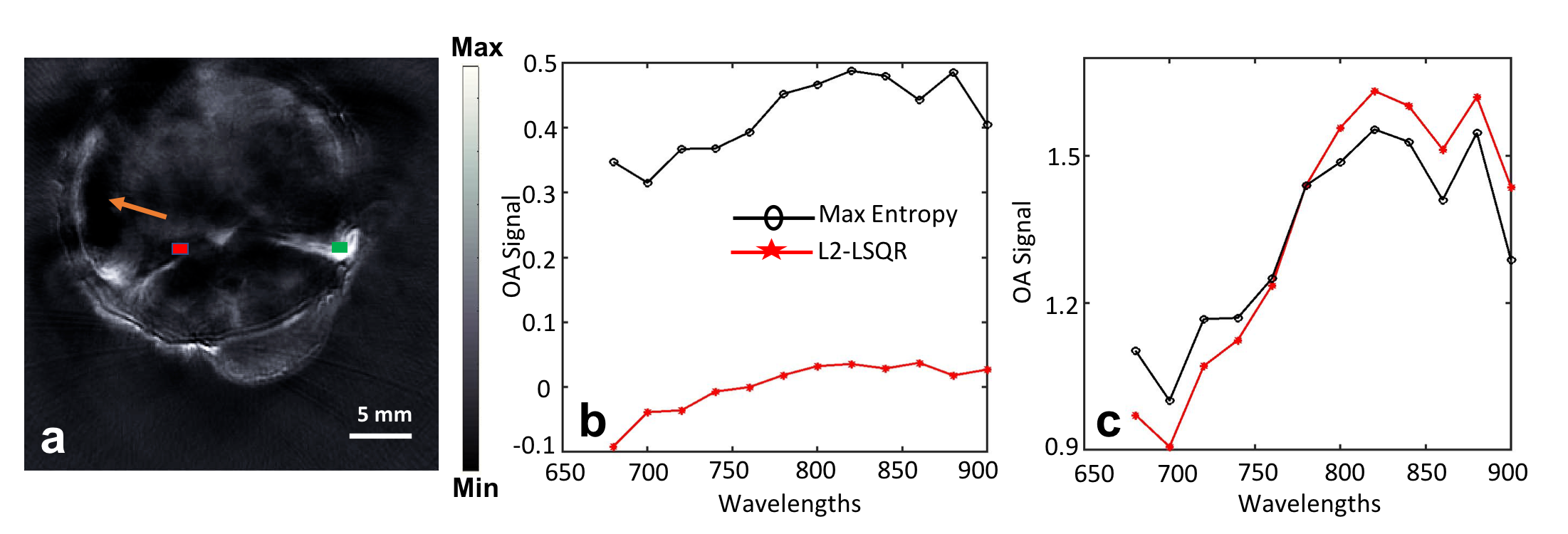}}
\caption{Comparison of entropy maximization scheme with standard L2-norm based reconstruction in terms of accurate spectral recovery. (a) Optoacoustic reconstruction using $\ell_2$-norm based reconstruction with thresholding, (b) Mean spectra shown for the region shown using red block in (a), (c) Mean spectra shown for the region shown using green block in (a); An 8 week old nude mice (CD-1 Nude, Charles River Laboratories, Germany) bearing a subcutaneous tumor was imaged at an wavelengths of 680 nm to 900 nm in steps of 20 nm.}
\label{fig7}
\end{figure}

Finally, we performed a study to check if entropy maximization scheme was able to accurately recover the spectral information. Fig. \ref{fig7}(a) shows the reconstruction results pertaining to a tumor bearing mice using L2-norm based scheme with thresholding at 680 nm wavelength. Fig. \ref{fig7}(b) shows the recovered mean spectral information using entropy maximization and L2-norm based reconstruction for the red square region shown in  Fig. \ref{fig7}(a).  Fig. \ref{fig7}(b) indicates that at wavelengths below 700 nm, we have appearance of negative values using L2-norm based reconstruction. Moreover, in some parts of the image, like the one shown using orange arrow in Fig. \ref{fig7}(a), the entire recovered spectra turned out to be negative using L2-norm based reconstruction (however maximum entropy scheme was able to recover positive spectral profile). Fig. \ref{fig7}(c) shows reconstructed mean spectra information using entropy maximization and L2-norm based reconstruction from the green square region indicated in Fig. \ref{fig7}(a). As can be seen from Figs \ref{fig7}(b) and \ref{fig7}(c), the spectral recovery of maximum entropy scheme is similar to that of L2-norm based reconstruction, however the appearance of negative values in L2-norm based reconstruction will hinder unmixing results in terms of absolute quantification.

\section{Discussion and Conclusion}
\label{discussion}
The reconstruction results for the numerical simulations, phantom and \textit{in-vivo} mouse scans indicate that the proposed entropy maximization scheme renders strictly positive image values that are also close to the a-priori known absorption values in the phantom. Employing a segmented image prior can effectively reduce the aberrations in image contrast by suitably mapping the light propagation pathway in two optically diverse domains (background and tissue), and enhance the performance of (optical) fluence correction methods\cite{b37}, as demonstrated in Figs 2(f) and 6(g).  Moreover, when a global SoS is attribute to the entire imaging domain, small SoS variation causes aberration at the edge of the surfaces of the imaged object \cite{b39}, the same two compartment model can be used to remove SoS mismatch. The figure of merits (Table-\ref{tab1}), magnitude of Fourier spectrum from the reconstructed images, and the line plots indicate entropy maximization approach provides superior results in comparison with non-negativity constrained reconstructions. Importantly the proposed approach offers an opportunity for exploring a family of differential type non-negative regularization methods (like entropy scheme).
\begin{table}
\caption{Evaluation of the methods:  Number of non-negative pixels and sharpness metric with the $\ell_2$-norm with non-negativity constraint and proposed maximum entropy method on different datasets.}
\setlength{\tabcolsep}{1pt}
\centering
\begin{tabular}{|p{65pt}|p{99pt}|p{65pt}|}
\hline
Metrics& 
No. of Non-Negative Values& 
Sharpness Metric\\
\hline \hline
\textbf{Star Phantom} & 
\makecell{$\ell_2$-NN = 11963 \\ MaxEn = 16890}& 
\makecell{$\ell_2$-NN = 0.0075 \\ MaxEn = 0.0125 } \\
\hline
\textbf{Murine Brain}& 
\makecell{$\ell_2$-NN = 7587 \\ MaxEn = 10741 }& 
\makecell{$\ell_2$-NN = 0.0121 \\ MaxEn = 0.0171}\\
\hline
\textbf{Murine Kidney}& 
\makecell{$\ell_2$-NN = 8224 \\ MaxEn = 15071}& 
\makecell{$\ell_2$-NN = 0.0092 \\ MaxEn = 0.0226} \\
\hline 
\multicolumn{3}{p{251pt}}{$\ell_2$-NN: $\ell_2$-norm Non-Negativity}\\
\multicolumn{3}{p{251pt}}{MaxEn: Maximum Entropy}
\end{tabular}
\label{tab1}
\end{table}

The entropy maximization scheme performed better with experimental data (Figs 2 and 6) compared to numerical simulation (Fig. 1). This is because experimental OA measurements are heavily influenced by experimental factors like laser pulse width, transducer impulse response, pitch and size of the transducer, making the reconstruction problem with experimental OA measurements more challenging. From Figs 2 and 6, it can be observed that the presence of negative pixels is higher in water region and in the center of imaging domain, where the absorption/the fluence is low resulting in lower SNR in time-series OA measurements. Similarly, introduction of noise and fluence effects in simulation studies (Fig. 1) results in large number of negative values in regions where the initial pressure rise is close to 0 and also generating spurious negative values inside the numerical breast phantom.

In recent studies, lot of emphasis has been placed on using $\ell_1$-norm based minimizations for performing OA tomographic image reconstruction in different frameworks \cite{b8, b12, b40}. We have performed $\ell_1$-norm based reconstruction as explained in \cite{b41} and the results pertaining to non-negativity constraint in the $\ell_1$-norm minimization is shown in Fig. S2. Fig. S2 also shows the performance comparison of $\ell_1$-norm minimization with entropy maximization and Tikhonov reconstruction with printed phantom data. We observe that applying a $\ell_1$-norm constraint does not afflict the appearance of negative values and the reconstruction performance is similar to $\ell_2$-norm based scheme in terms of reducing negative values. This also demonstrates the superiority of using entropy maximization to generate physically relevant OA reconstructions devoid of negative values. We have not taken up further comparisons with $\ell_1$-norm based approach, as our goal was to demonstrate the utility of entropy maximization approach to overcome appearance of pixels with negative values.

Entropy maximization scheme was evaluated with biological datasets acquired from $270^{\circ}$ detection angle wherein the acquired dataset consists of highly independent (incoherent) data. While recent developments involve building systems with handheld probes ($90^{\circ}$ three-dimensional acquisition, or $145^{\circ}$ two-dimensional acquisition) with different data-collection geometry. Performing accurate reconstructions with these clinical handheld systems tend to be difficult due to acquisition of limited independent data. Evaluating the performance of the entropy scheme with the limited independent data scenarios can enable utility of OA imaging in different clinical scenarios\cite{b42_1}. 

The proposed method preserves the structural integrity (numerical breast phantom and star phantom) and the anatomical structures (mouse data), and was successful in correcting the effects of variations in optical fluence. As part of future work, we aim to integrate the entropy maximization with more accurate light propagation modeling (such as Monte Carlo based schemes) to obtain better representation of the absorption coefficient with the reconstruction process accelerated by means of graphics processing units \cite{b42}. In this work, we demonstrated a non-negative image reconstruction method with improved image quality using fluence correction step at single acquisition wavelength. Translating the same to multi-wavelength scenario for estimation of quantitative tissue parameters is a fairly complex problem, since the optical properties used for fluence estimation varies nonlinearly with wavelength and is not known beforehand. Combining these problems will lead to generation of infinite possible ways to obtain accurate spatio-spectral representation, and such spectral analysis methods are beyond the scope of the current study. 

In this work, we have shown that entropy maximization is able to accurately recover the spectral information compared to L2-norm based reconstruction (see Fig. 7). However, the ability to resolve intrinsic chromophores like oxyhemoglobin, deoxyhemoglobin, fat, and water by acquiring data at multiple wavelengths is a key benefit of multispectral OA imaging. The unmixing of chromophores is achieved by a solving system of linear equations (direct or non-negatively constrained), or by non-linear unmixing using an integrated fluence correction. All of these approaches use thresholding of negative values, making them suboptimal and error prone. On the other hand, entropy maximization can  purge out the inaccuracies occurring from truncated pixel information, potentially improving the performance of unmixing and image analysis algorithms. Therefore, the future work will involve comparing the different combination of reconstruction (acoustic inverse problem) and unmixing with different solvers like LSQR, non-negative LSQR and entropy maximization to bring out value among these schemes.

\section{Conclusion}
\label{sec:conclusion}
The proposed maximum entropy based OA image reconstruction scheme demonstrates superior reconstruction performance with no visible distortion of anatomical structures associated with delivering of non-negative pixel values. Entropy maximization reconstruction thus tends to be physically relevant and more accurate in resolving the structures (as demonstrated with numerical simulation, experimental phantoms and \textit{in-vivo} case) in an imaged sample. The developed methodology has the potential to emerge as a suitable data processing tool for OA imaging, and specifically benefiting pre-clinical biomedical \cite{b43} and translational imaging \cite{b44}.
 
\appendices

\section{Implicit Non-negativity using Entropy Maximization}

The objective function in the entropy maximization scheme is given as,
\begin{equation}
 {\rm{\Omega }} = \left| {\left| {{\bf A}x - b} \right|} \right|_2^2 + \lambda {x^T}{\rm{\;}}\log \left( {\frac{x}{m}} \right)
\end{equation}
The gradient of the above equation can be written as,
\begin{equation}
\frac{{\partial {\rm{\Omega }}}}{{\partial x}} = {{\bf A}^T}\left( {{\bf A}x - b} \right) + \lambda \left( {1 + \log \left( {\frac{x}{m}} \right)} \right) = 0
\end{equation}
Now, we can consider the above minimization problem as minimizing two models in the subspace, one is based on residual i.e. $Res = \left| {\left| {Ax - b} \right|} \right|_2^2$ and the other being relative entropy i.e. $Ent = \sum x\log \left( {\frac{x}{m}} \right)$. Here the regularization parameter defines the proportion of residual and entropy term in this minimization problem. As in any optimization, the solution is always found using the search directions (these search directions are defined by the gradients). The update equation at $i^{th}$ gradient iteration will turn out to be,
\begin{equation}
{x_i} = {x_{i - 1}} - \alpha {\left( {\frac{{\partial {\rm{\Omega }}}}{{\partial x}}} \right)_{{x_{i - 1}}}}    
\end{equation}
where $\alpha$ is the step length estimated using line search method and is always non-negative. As ${x_{i - 1}} \to 0,{\rm{\;}}\nabla Ent \to  - \infty {\rm{\;}}$, the gradient update will be pushed to a very low value using entropy constraint. Also note that as, ${x_{i - 1}} \to 0$, $\nabla Ent$ will reach $ - \infty $ faster, and the $\nabla Res \to  - {A^T}b$; importantly $\nabla Res$ cannot reach $\infty$ as fast as $\nabla Ent \to - \infty$ to nullify the effect of entropy term, therefore the overall gradient will be negative i.e. ${\left( {\frac{{\partial {\rm{\Omega }}}}{{\partial x}}} \right)_{{x_{i - 1}}}} \to  - ve$. In any gradient descent method, we traverse in the direction perpendicular to the gradient, therefore the solution will be pushed away from zero to have high positive value, i.e. as ${x_{i - 1}} \to 0,{\rm{\;}}{x_i} \to  + ve$.  Hence, using the entropy constraint will enable the solution to move away from zero and leading to positive real numbers. Since, a natural barrier is created by including the entropy constraint into the optimization framework, this barrier will not allow the solution to take negative values and consequently positive OA reconstructions are generated. In order to converge to positive OA reconstructions, we need to start with a large positive initial guess i.e. when ${x_0} \to {\rm I\!R{^ +}}$ then $\nabla Res \to {A^T}A{x_0}$ and $\nabla Ent \to {\rm I\!R{^ +}}$. Further, using a step-length control i.e. $\alpha  = \min {( - \nabla {\rm{\Omega }}\left( {{x_{i - 1}}} \right))^T}\nabla {\rm{\Omega }}\left( {{x_{i - 1}} - \alpha \nabla {\rm{\Omega }}\left( {{x_{i - 1}}} \right)} \right)$ will ensure positive OA reconstructions, because the choice of $\alpha$ (estimated using secant method) would ensure positive solution in next iteration ${x_i} = {x_{i - 1}} - \alpha \nabla {\rm{\Omega }}\left( {{x_{i - 1}}} \right)$.

\end{document}